\documentclass[twocolumn,amsmath,amssymb,prx]{revtex4}
\usepackage{latexsym}
\usepackage{graphicx}
\usepackage{color}
\usepackage{amsmath}
\newcommand{\beginsupplement}{%
        \setcounter{table}{0}
        \renewcommand{\thetable}{S\arabic{table}}%
        \setcounter{figure}{0}
        \renewcommand{\thefigure}{S\arabic{figure}}%
     }

\begin{document}
\title{Gate Controlled Anomalous Phase Shift in Al/InAs Josephson Junctions}

\author{William~Mayer$^{1}$}
\author{Matthieu C. Dartiailh$^{1}$}
\author{Joseph~Yuan$^{1}$}
\author{Kaushini~S.~Wickramasinghe$^{1}$}
\author{Enrico~Rossi$^{2}$}
\author{Javad~Shabani$^{1}$}
\affiliation{$^{1}$Center for Quantum Phenomena, Department of Physics, New York University, NY 10003, USA\\
$^{2}$ Department of Physics, William \& Mary, Williamsburg, VA 23187, USA
}

\date{\today}

\maketitle


\textbf{In a standard Josephson junction the current is zero when the phase difference between the superconducting leads is zero. This condition is protected by parity and time-reversal symmetries. However, the combined presence of spin-orbit coupling and magnetic field breaks these symmetries \cite{Rasmussen:2016} and can lead to a finite supercurrent even when the phase difference is zero \cite{Shen:2014,Konschelle:2015}. This is the so called anomalous Josephson effect --  the hallmark effect of superconducting spintronics --and can be characterized by the corresponding anomalous phase shift ($\phi_0$) \cite{Szombati:2016,Assouline:2019}. We report the observation of a tunable anomalous Josephson effect in InAs/Al Josephson junctions measured via a superconducting quantum interference device (SQUID). By gate controlling the density of InAs  we are able to tune the spin-orbit coupling of the Josephson junction by more than one order of magnitude. This gives us the ability to tune $\phi_0$, and opens several new opportunities for superconducting spintronics \cite{Linder:2015}, and new possibilities for realizing and characterizing topological superconductivity \cite{Alicea:2012,Fornieri:2019,Ren:2019}.}

Superconductivity and magnetism have long been two of the main focuses of condensed matter physics. Interfacing materials with these two opposed types of electron order can serve as a platform to host many new phenomena. Recently these systems have drawn renewed theoretical and experimental attention in the context of superconducting spintronics \cite{Linder:2015}  and in the search for Majorana fermions \cite{FuKane2008,Elliott:2015,Lutchyn:2010,Oreg:2010}. Novel heterostructures can provide the ingredients that are typically needed: superconducting pairing, breaking of
time reversal symmetry, and strong spin-orbit coupling. 

A basic property of superconducting systems is that we can introduce a relation between charge current and the superconductor's phase. In the canonical example of a Josephson junction (JJ), this is the current-phase relationship (CPR), where $\phi$ is the phase difference between the two superconductors. Systems with nontrivial spin texture generally introduce a relationship between charge and spin. In the case of spin-orbit coupling this can manifest in many ways including the spin Hall effect and topological edge states \cite{Awschalom:1174}.

A hybrid system, combining spin-orbit coupling and superconductivity, results in a much richer physics where phase, charge current and spin are all interdependent. This gives rise to new phenomena such as an anomalous phase shift which is the hallmark effect of superconducting spintronics~\cite{Linder:2015}. In a standard JJ, the CPR always satisfies the condition $I(\phi=0)=0$. This condition is protected by parity and time-reversal symmetries. However the presence of spin-orbit coupling along with the application of an in-plane magnetic field can break these symmetries \cite{Rasmussen:2016}. This allows an anomalous phase ($\phi_0$), which means that with no current flowing there can be a non-zero phase across the junction or, conversely, at zero phase a current can flow. This is also understood in the context of the spin-galvanic effect, also known as the inverse Edelstein effect. It states that in a normal system with Rashba spin-orbit coupling, a steady state spin gradient can generate a charge current \cite{Shen:2014}. When superconductivity is introduced, gauge invariance no longer prohibits a finite static current-spin response \cite{Konschelle:2015}. Consequently in the superconducting state, a static Zeeman field can induce a supercurrent, which can be measured as $\phi_0$.

Anomalous phase junctions were demonstrated in InSb nanowires in a quantum dot geometry \cite{Szombati:2016} and more recently in JJ using Bi$_{2}$Se$_{3}$ \cite{Assouline:2019}. In the quantum dot realization the phase shift is gate tunable but is geometrically constrained and  only supports a few modes and consequently small critical currents. In Bi$_{2}$Se$_{3}$, a topological insulators, large planar $\phi_0$-junction are possible, however Bi$_{2}$Se$_{3}$ is not gate-tunable.

Our work is based on heterostructures formed by InAs and epitaxial superconducting Al \cite{Shabani:2016} which have emerged as promising heterostructures not only for mesoscopic supercondcutivity \cite{Bottcher:2018} but also for the realization of topological superconductivity and Majorana fermions \cite{Suominen:2017}. This is due to the fact that the induced superconducting gap, $\Delta_{\rm ind}$, in InAs can be as large as the one in Al~\cite{Kjaergaard:2016}, and InAs has large g-factor and spin orbit coupling. As a consequence, JJ fabricated on this platform can have large critical current and high transparency \cite{Kjaergaard2017, Mayer:2019}. Furthermore, one can control the strength of the spin-orbit coupling by tuning the density in the InAs via external gates \cite{Kaushini2018}.

\begin{figure}[ht]
\centering
\includegraphics[width=0.48\textwidth]{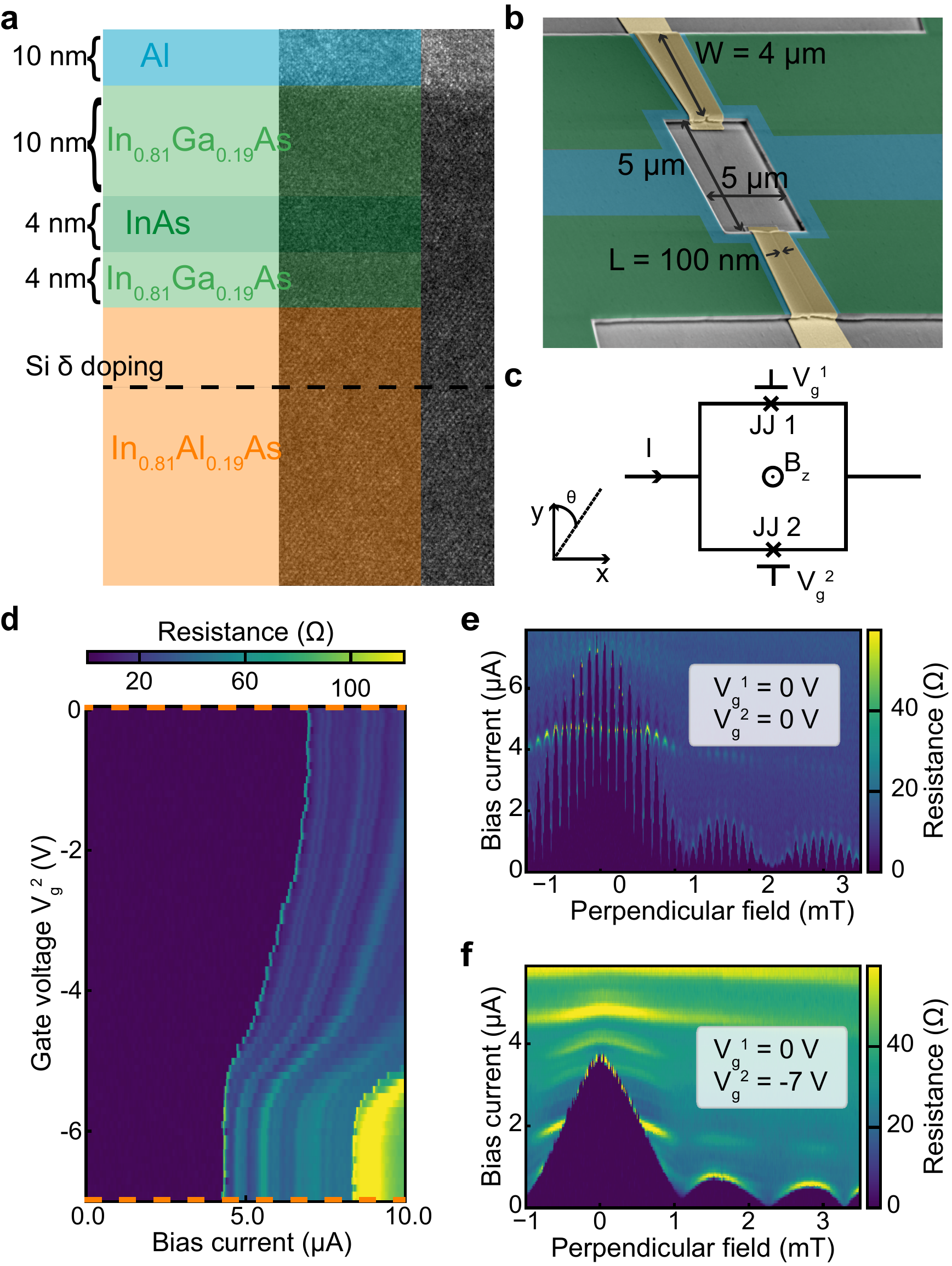}
\caption{\label{figure1}(Color online) 
(a) Sample stack description super-imposed on large scale TEM image. 
(b) Colorized SEM image of a device similar to the one presented. The SQUID loop is about 5$\mu$m$\times$5$\mu$m, both junctions have a gap of about 100 nm and are 4 $\mu$m wide. 
(c) Schematic of the device. Each junction can be gated independently. The x direction is defined in the plane of the sample along the current direction. 
(d) Resistance of the SQUID as a function of the applied bias current and the gate voltage applied on JJ 2. The dashed orange line at the top and bottom of the graph indicates the gates parameters used in (e) and (f) respectively.
(e) Resistance of the SQUID as a function of the perpendicular field and bias current with both gates set at 0 V. Typical fast SQUID oscillation of the critical current can be seen on top of the larger scale Fraunhofer pattern of the junctions.
(f) Resistance of the SQUID as a function of the perpendicular field and bias current with $V_g^1$ set at 0 V and $V_g^2$ set at -7 V. The SQUID oscillations visible in (e) are completely absent and only the single junction Fraunhofer pattern is visible.}
\end{figure}

Figure~1a shows a transmission electron microscope image of the heterostructure with false colors. We fabricate SQUID loops consisting of two Al/InAs JJ's in parallel. The fabrication details were previously reported \cite{Kjaergaard:2016} and are detailed in Methods. Figure~1b shows a tilted view scanning electron microscope image of a device with false colors, and the device schematic is depicted in Fig. 1c. Both junctions are 4 $\mu$m wide (W) and 100 nm long (L) while the size of the SQUID loop is 25 $\mu m^2$. The high aspect ratio of the junction ($W/L$) yields devices that have many transverse modes and consequently large critical currents. Typical mean free path ($l_e$) in the semiconductor region is near $l_e \simeq 200$~nm and the superconducting coherence length ($\xi$) is estimated to be $\xi=770$nm \cite{Mayer:2019}. The two junctions show small variations in normal resistance($R_n$), $R_{n}^1 = 102 \:\Omega$, $R_{n}^2 = 110\:\Omega$ and critical current($I_c$) $I_{c}^1 = 4.4 \:\mu A$, $I_{c}^2 = 3.6 \:\mu $A when gates are not activated. Gate voltage ($V_g$) varies the density of the InAs region thereby changing $R_n$ and $I_c$ of each JJ. 

Figure~1d shows the variation in critical current while changing only $V_{g}^2$. At low voltages, $V_g^2 < -5.5\:\textrm{V}$, the critical current becomes constant indicating that JJ2 is fully depleted. This is confirmed by phase bias measurements performed by applying perpendicular magnetic field ($B_z$), shown in Fig.~1e \& 1f. In Fig.~1e, when both junctions are at $V_g^{1} = V_g^{2} = 0\:\textrm{V}$, we see characteristic SQUID oscillations with application of $B_z$. Superimposed on top of the fast SQUID oscillations is the much slower Fraunhofer diffraction pattern from each individual JJ. Conversely when $V_{g}^2 =-7\:\textrm{V}$, in Fig. 1f, we observe only the Fraunhofer pattern indicating the presence of only a single JJ. This allows us to effectively study each JJ individually. 

 Individual JJs are characterized in in-plane magnetic field. We find $B_c = 1.45\:\textrm{T}$ for thin film Al in both junctions and is independent of the in-plane field direction. However, $I_c$ of both JJs show a strong asymmetry in in-plane magnetic field. We observe a stronger decrease in $I_c$ as a function of $B_x$(field applied along the current direction). This is consistent with previous measurements on InAs 2DEG based JJ \cite{Suominen:2017}, and recent work suggests this could be related to the nature of spin-orbit coupling in the system \cite{Bommer:2019}. Measurements of Fraunhofer pattern with increasing in-plane field show increasing asymmetry, but unlike previous studies this asymmetry is found to be independent of in-plane field direction. Additionally, despite the distortions, the Fraunhofer pattern appears to remain periodic indicating a homogeneous current distribution at all fields. Figures and further discussion are presented in Supplemental.

\begin{figure}[ht]
\centering
\includegraphics[width=0.48\textwidth]{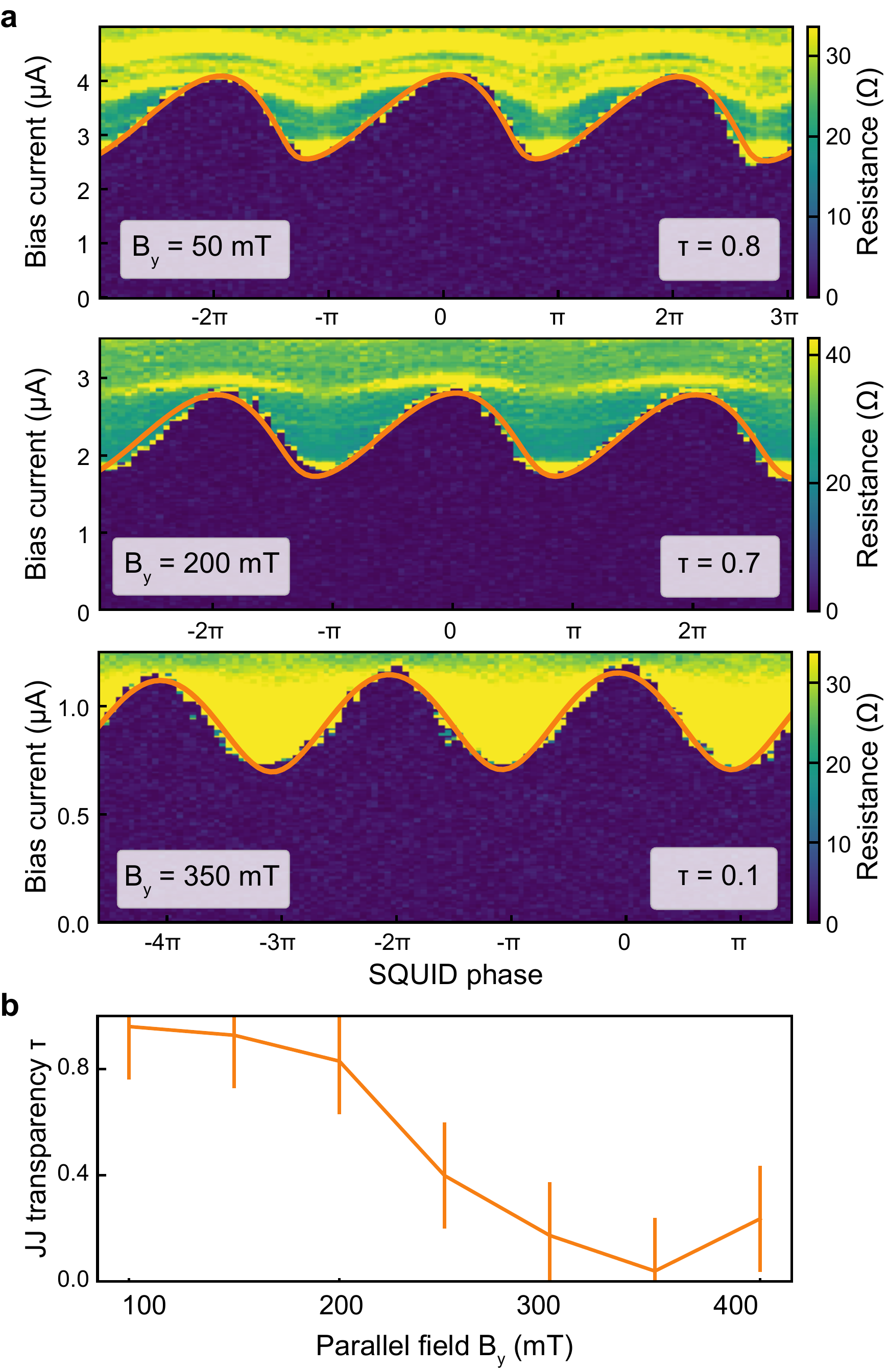}
\caption{\label{figure2}(Color online) (a) Resistance of the device as a function of the phase bias applied on the SQUID and the bias current in the presence of an in-plane field along the y direction at $B_y = 50\:\textrm{mT}$, $B_y = 200\: \textrm{mT}$ and $B_y = 350 \:\textrm{mT}$. $V_g^1$ is set to -2 V and $V_g^2$ to -4.5 V resulting in $I_c^1\approx4I_c^2$. The oscillation of the critical current present a visible forward tilt at 50 and 200 mT absent at 350 mT.
(c) Evolution of the transparency of JJ2 as a function of the in-plane field By as determined from fitting the SQUID oscillation at different gate and fields (see Methods)
}
\end{figure}

Measurements of robust Fraunhofer pattern up to $B_y=400\:\textrm{mT}$ are made possible in this system due to the large induced gap in the semiconductor region \cite{Kjaergaard2017}. Using the product $I_cR_n/\Delta$, where $\Delta$ is the superconducting gap of the Al, the quality of the junction can be characterized. For the junctions used in this study we measure $I_{c}^1R_{n}^1/\Delta =2$ \& $I_{c}^2R_{n}^2/\Delta =1.78$. Studies of CPR can also aid junction characterization, as a non-sinusoidal CPR indicates a highly transparent JJ. Measurements of skewed CPR have been demonstrated in InAs nanowire JJ \cite{Konschelle:2015} bismuth nanowires \cite{Murani:2017} and graphene devices \cite{Thompson:2017}. The generalized CPR can be described by Eq. 1, where $\phi_t$ is the total phase across the junction, $\tau$ is the junctions transparency and we neglect any temperature dependence since all measurements are performed at $\textrm{T}=20\:\textrm{mK}$:

\begin{equation}
    I(\phi_t) = I_c\,\frac{\sin{\phi_t}}{\sqrt{1 - \tau\sin^2{\phi_t/2}}}
\end{equation}

To measure the CPR, we apply gate voltages to the junctions to create a highly asymmetric current configuration($I_c^1\approx4I_c^2$). This effectively fixes the phase of the high current junction so we measure only the CPR of the lower current junction. Figure~2a shows resistance maps at $B_y=50\:\textrm{mT}$, $B_y=200\:\textrm{mT}$ and $B_y=350\:\textrm{mT}$ in the CPR regime. At $B_y=50\:\textrm{mT}$ the plot shows a forward skew indicating high JJ transparency. To fit the SQUID oscillations, we sum the contributions of each JJ with a phase difference due to applied $B_z$ and maximize the current with respect to the sum of the phases. The resulting fits are shown in Fig. 2a as orange overlays. The transparencies obtained from the fits are indicated in each plot.  Measurement at $B_y=350\:\textrm{mT}$ reveals the oscillations are more sinusoidal, indicating reduced transparency. The dependence of transparency on $B_y$ for JJ2 is shown in Fig.~2b. We observe near unity transparency at low fields with a rapid decline above $200\:\textrm{mT}$. Both junctions show similar dependence of transparency on $B_y$. The mechanism leading to the decreased transparency as a function of $B_{y}$ is not well understood. Note that these fits are based on the assumption that the JJ CPR is captured by Eq.~1. 

\begin{figure*}[ht]
\centering
\includegraphics[width=0.98\textwidth]{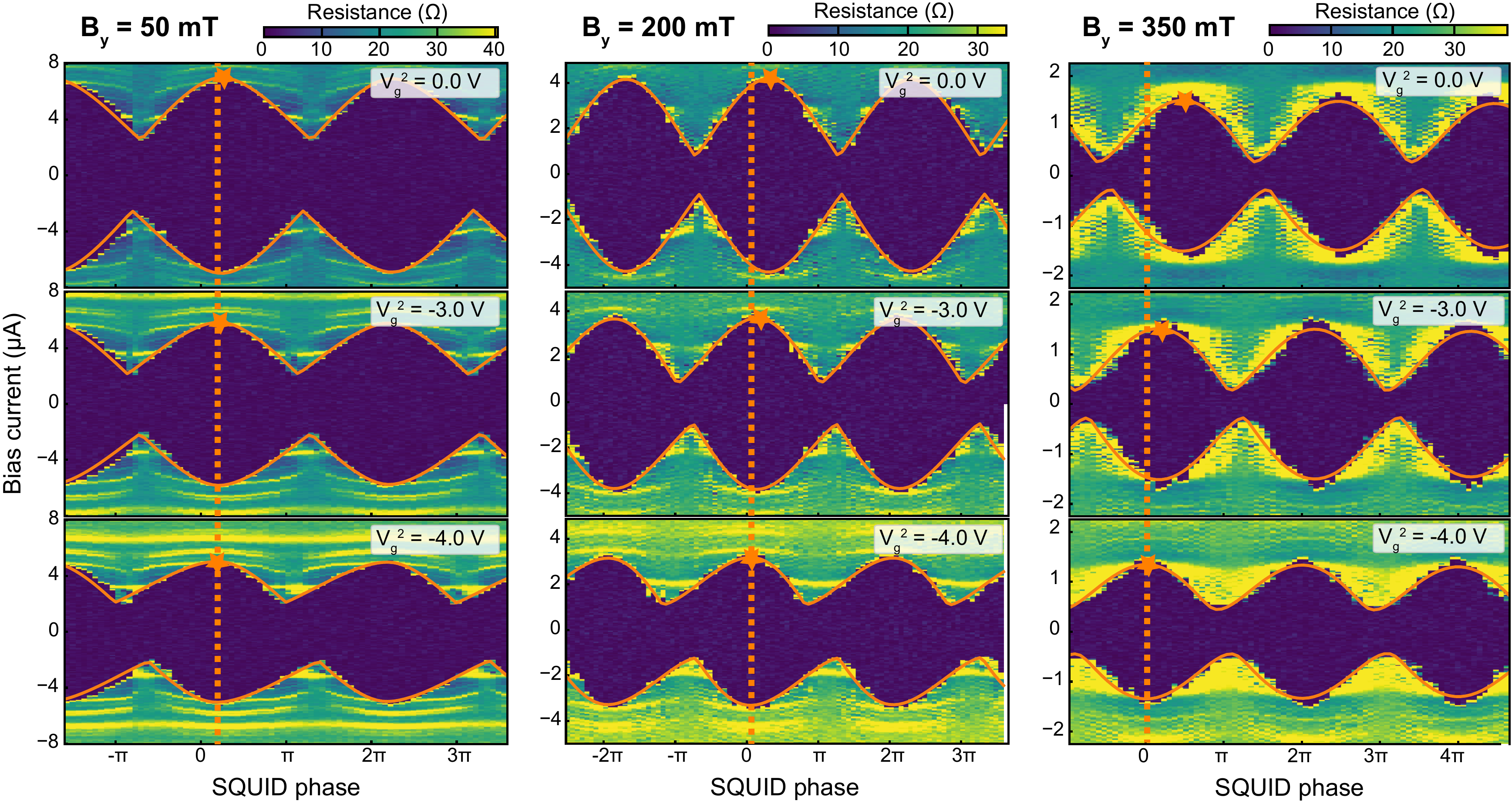}
\caption{\label{figure3}(Color online)  Resistance of the device as a function of the phase bias applied on the SQUID and the bias current at three different value of the in-plane field By and three different values of $V_g^2$. In all scans $V_g^1$ is set to -2 V. The dashed orange line indicates the position of the maximum of the oscillation at $V_g^2 = -4 \:\textrm{V}$. The orange stars indicates the position of the maximum at each field. }
\end{figure*}

If we consider a single JJ with an anomalous phase a typical current biased measurement will show no measurable signature. When a JJ is current-biased, the CPR dictates that the phase will change so the critical current is maximized. This means that any phase shift applied to such a system will be invisible once current is maximized. A simple alternative which has been employed in previous studies of $\phi_0$ is to use a SQUID geometry, whose primary property is phase sensitivity. Even in a SQUID, any single scan generally has an phase offset obscuring the effect of $\phi_0$. In order to experimentally measure $\phi_0$ a phase reference is necessary. To this end we compare scans taken consecutively at the same field but changing $V_g$ of one JJ. The gate voltage varies both the density and strength of spin-orbit coupling which should change $\phi_0$. Figure~3 shows resistance maps taken at different $B_y$ for three $V_g^2$. By finding the phase shift between these different gate voltages we can measure the variation of $\phi_0$. This shift is most easily seen by comparing the positions of SQUID oscillation maxima at different $V_g^2$. To extract the phase difference we fit the data using a similar procedure as applied to the CPR of Fig.~2. The only adjustment is now we include $\phi_t=\phi+\phi_0$ in each CPR relation. In the case of a varying transparency, one could observe an apparent phase-shift unrelated to $\phi_0$. However this shift would have the opposite sign on the positive and negative bias branches of the measurement. The data presented in Fig. 3 are symmetric in bias, which allows us to definitively separate the effects of transparency and a $\phi_0$ shift. A more detailed description of the fitting can be found in Methods.

The anomalous phase $\phi_0$ is expected to grow with the strength of the spin-orbit coupling. Previous work on InAs indicates that the Rashba spin-orbit coupling can be tuned from close to zero to as high as $180\:\textrm{meV}\AA$, with apparent saturation at high densities \cite{Kaushini2018}. This indicates that $\phi_0$ should be smallest at the lowest gate voltages. Consequently, we take $V_g=-4\:\textrm{V}$ as the reference scan which allows us to minimize the reference contribution to $\Delta\phi_0$, i.e., the difference $\phi_0(V_g)-\phi_0(-4V)$. Figure~4a shows how $\Delta\phi_0$, extracted from the fits, increases with gate voltage and saturates at higher $V_g$. In \cite{Kaushini2018} it was shown that $\alpha$ increases as density (n) increases but that for low densities the relationship is non-linear. This could explain the general $V_g$ dependence of $\phi_0$ since at low $V_g$ $\alpha$ is increasing faster than $n$ leading to a rapid increase of $\phi_0$ versus $V_g$ , while at higher $V_g$ the effect of $\alpha$ and $n$ cancel out and the $\phi_0$ dependence on $V_g$ weakens.



\begin{figure}[ht]
\centering
\includegraphics[width=0.48\textwidth]{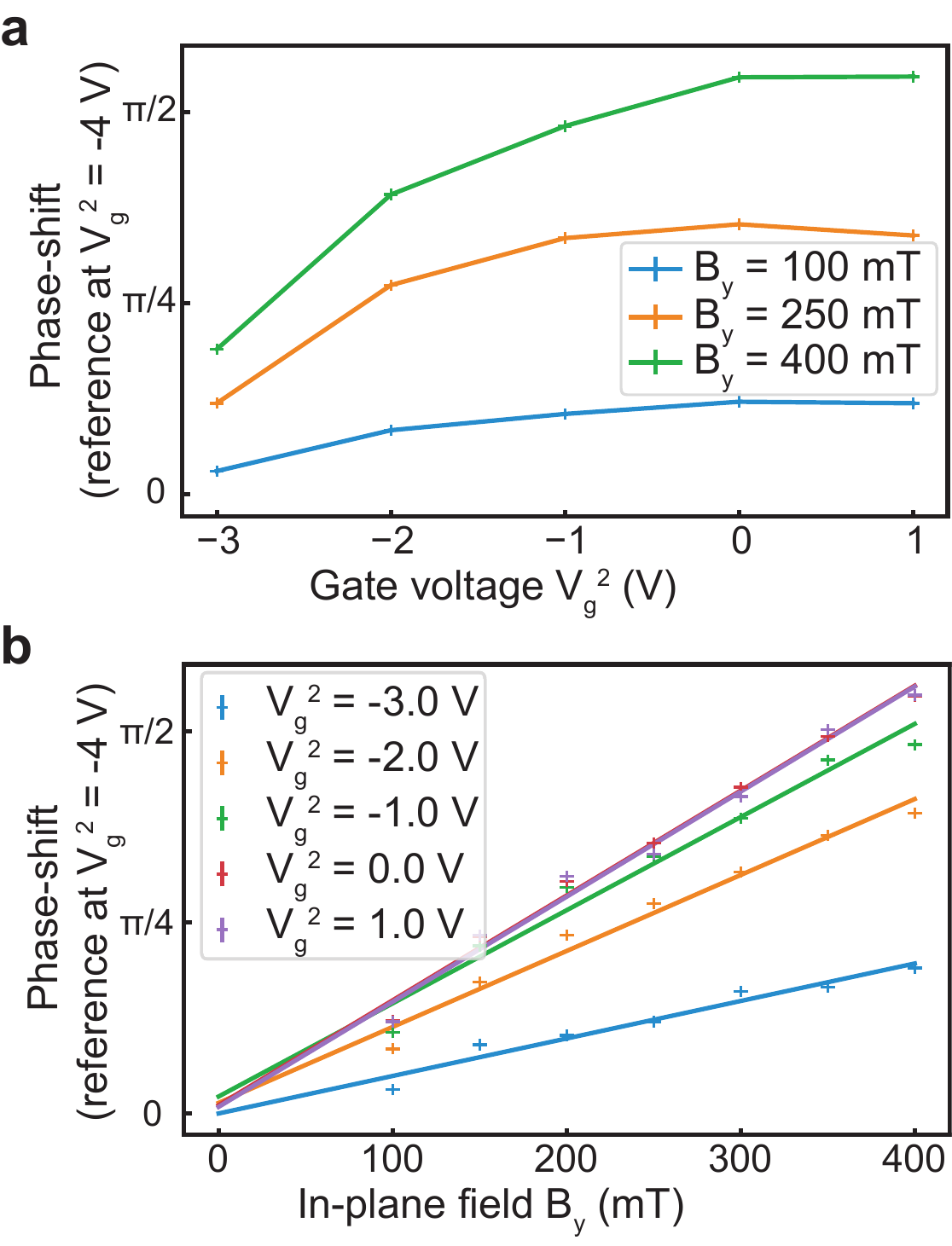}
\caption{\label{figure4}(Color online) Evolution of the phase-shift in JJ2 as a function of the gate voltage (a) and of the applied in-plane field along y (b). The phase-shift $\Delta\phi_0$ is measured between the oscillations at a given value of $V_g^2$ and the ones at -4 V used as reference. In (b) the solid lines corresponds to linear fits to the measured phase-shifts.}
\end{figure}

Several theoretical works have studied the interplay of spin-orbit coupling and time reversal breaking fields in JJs. They provided scalings of $\phi_0$ with respect to material and geometry parameters~\cite{Krive2004,Buzdin:2008,Yokoyama2014a,Tokatly:2015}. Almost all the available theoretical works consider the long junction limit in which the distance $L$ between the superconductors is much larger than the coherence length $\xi$. In this limit theory predicts, for a single transverse mode, $\phi_0=4\alpha L\,E_z/(\hbar v_F)^2$ in the ballistic regime~\cite{Buzdin:2008}, and $\phi_0={m^*}^{2}(\alpha L)^3\,E_z/(\hbar^3v_F)^2$ in the diffusive regime~\cite{Tokatly:2015}, where $m^*$ is the effective mass and $v_F$ is the Fermi velocity. 

Both analytic expressions reflect the fact that the anomalous Josephson effect is expected to be stronger as the ratio $L/\xi$ increases. However, by substituting in these expressions our material parameters we find that both results return values of $\phi_0$ that are much smaller than the ones we observe.
This is not surprising considering that in our devices $\xi\sim 770\:\textrm{nm}$. In addition, the disorder is quite low resulting in a mean free path $l_{\rm e}\simeq 200\:\textrm{nm}$, so that results obtained in the limit $L\gg\xi$ are not directly applicable to our system.

The large value of $\phi_0$ can be qualitatively understood considering that in our devices we have a very large number of transverse modes. For a few of these modes $v_F$ is very small and therefore $L/\xi>1$. Disorder and the spin-orbit coupling term $k_y\sigma_x$ can both mix the spin-split transverse modes resulting in the breaking of the chiral symmetry of the Andreev bound states, and therefore significantly contribute to $\phi_0$.

Figure~4b shows the dependence of $\Delta\phi_0$ on $B_y$ at a range of gate voltages. The strong agreement with linear fits confirms that $\Delta\phi_0$ is proportional to the Zeeman energy in agreement with theory \cite{Konschelle:2015}. With a more complete theoretical understanding in the limit of strong proximity effect it should be possible to estimate the strength of spin-orbit coupling from the slope of the anomalous phase dependence. At the largest $B_y$ and $V_g$ measured we observe $\Delta\phi_0>\pi/2$ setting a lower bound on $\phi_0$. It is possible to optimize both L and W of each JJ to increase $\Delta\phi_0$, and consequently $\phi_0$.

In summary, we have shown the capability to tune the
anomalous phase shift of Josephson Junctions formed by InAs and Al. This tunability results from the ability to vary the strength of the spin-orbit coupling via an external gate. The observation of a finite $\phi_0$ indicates a coupling of the superconductors phase, charge current, and spin in these heterostructures. We find $\phi_0$ to be proportional to the Zeeman energy, as expected, and its magnitude to be much larger than the currently available  theoretical scalings. This is most likely due to the fact that such scalings are valid for in a long junction with few channels, a limit that is not directly relevant to our system. 

The capability to realize a large value of $\phi_0$, and to tune it, is of great importance for applications in superconducting spintronics where large spin gradients  can be used to realize phase batteries~\cite{Linder:2015}, and opens the possibility to generate, in a controllable way, spin gradients through Josephson currents or a phase bias. In addition, the observation that a significant $\phi_0$ can be present in InAs/Al heterostructures, and the fact that it strongly depends on the density of InAs, are directly relevant to efforts to realize topological superconducting states. In particular, the knowledge that an intrinsic phase difference $\phi_0$ can be present in InAs/Al Josephson junctions is of great importance  for recent proposals to realize topological superconductivity in phase-controlled Josephson junctions~\cite{Ren:2019,Fornieri:2019}.

\vspace{.5cm}
This work is supported by DARPA Topological Excitations in
Electronics (TEE) program and NSF. We acknowledge fruitful discussions with Igor Zutic and Alex Matos-Abiague. ER acknowledges support from NSF, ONR and ARO, and helpful discussions with Joseph Cuozzo and Stuart Thomas. This work was performed in part at the Advanced Science Research Center NanoFabrication Facility of the Graduate Center at the City University of New York.

\section*{Methods}

\subsection{Growth and fabrication:}

The structure is grown on semi-insulating InP (100) substrate. This is followed by a graded buffer layer.
The quantum well consists of a 4 nm layer of InAs grown on a 4 nm layer of In$_{0.81}$Ga$_{0.25}$As and finally a 10 nm In$_{0.81}$Ga$_{0.25}$As layer on the InAs which has been found to produce an optimal interface while maintaining high 2DEG mobility \cite{Kaushini2018}. This is followed by $in-situ$ growth of epitaxial Al (111). Molecular beam epitaxy allows growth of thin films of Al where the in-plane critical field can exceed ~2~T \cite{Shabani:2016}. 

Devices are patterned by electron beam lithography using PMMA resist. Transene type D is used for wet etching of Al and a III-V wet etch ($H_2O:C_6H_8O_7:H_3PO_4:H_2O_2$) is used to define deep semiconductor mesas. We deposit 50 nm of $AlO_2$ using atomic layer deposition to isolate gate electrodes. Top gate electrodes consisting of 5 nm Ti and 70nm Au are deposited by electron beam deposition. 

\subsection{Measurements:}

All measurements are preformed in an Oxford dilution refrigerator with a base temperature of 7 mK. The system is equipped with a 6:3:1.5 T vector magnet. All transport measurements are performed using standard dc and lock-in techniques at low frequencies and excitation current $I_{ac}=10nA$. Measurements are taken in a current-biased configuration by measuring R=dV/dI with $I_{ac}$, while sweeping $I_{dc}$. This allows us to find the critical current at which the junction or SQUID switches from the superconducting to resistive state. It should be noted we directly measure the switching current, which due to effects of noise can be lower than the critical current. For the purposes of this study we assume they are equivalent.

\subsection{Fitting procedure:}

As illustrated in Fig. \ref{figure2}, the junctions forming the SQUID display a saw-tooth like current-phase relationship (CPR) characteristic of junctions with high transparencies, and this even at low gate. We hence model the CPR using Eq. 1 in which we neglect the temperature dependence which would only induce minor corrections. To model the SQUID pattern, we sum the contributions of two Josephson junctions with a phase difference and maximize (minimize for negative bias current) the current with respect to the sum of the phases. This requires the use of 6 parameters: the out-of-plane magnetic field to phase conversion factor, the transparency of each junction, the critical current of each junction (defined as independent of the transparency) and a phase. This represents a large number of parameters for fitting a single trace. To improve the accuracy of our procedure we consider multiple traces and reduce the number of parameters based on physical arguments.

Since we cannot experimentally access a reliable phase reference, we always compare measurements, taken within a single magnetic field sweep, for different values of the gate voltage applied to one of the junction (referred to as the active junction). The second junction (idler) stays at a constant gate voltage. We can hence fix the amplitude of the idler current for a given parallel field.

Changes in the transparency of a junction can cause an apparent phase-shift when considering only the positive bias current branch of the SQUID oscillation. However this apparent shift would have the opposite sign for the negative bias current branch. We have checked, as illustrated in Fig. \ref{figure3}, that the phase-shift we observe is present with the same sign on both branches. As a consequence we can reasonably assume that the transparency of the junctions is constant over the gate voltage range considered. This assumption allows us to use one transparency value per junction at a given field. The transparency value is better constrained in a CPR-like measurement and this is why to have a well constrained problem, we combine data sets taken in both configurations: JJ1 as active junction and JJ2 as idler and JJ2 as the active junction and JJ1 as idler.

Considering measurements at N parallel fields with M different gate values in both configuration (JJ1 active/JJ2 active), we fit for each junction N transparencies, N amplitudes as idler, N$\times$M amplitudes as active. Furthermore we extract 2$\times$N$\times$M phases. Because the field to phase conversion factor depends only on geometrical considerations we use a single value for each configuration \footnote{We observed that for data sets taken several weeks apart we could see small changes in the field to phase conversion factor, that we attribute to the magnet. As a consequence we use different factors for data taken when tuning JJ1 or JJ2.}. For the most extensive dataset, presented in Fig. 4, N = 7 and M = 6. Similarly, we can also take into account the Fraunhofer envelope of the oscillation using two global parameters: a period and a phase.

By comparing the transparencies from independent measurements of JJ1 and JJ2 at a given magnetic field, we find that the junction transparencies are very similar. Hence, the data for Fig. 2a, and Fig. 3 have been fitted using the equal transparencies assumption. The data for Fig. 2b and 4 have been fitted using the full method presented above but we focused on JJ2 results.

\clearpage

\section*{Supplementary Material}
\beginsupplement

\title{Supplementary materials: Gate Controlled Anomalous Phase Shift in Al-InAs Josephson Junctions}

\section{Fraunhofer pattern in the presence of a parallel field}

\begin{figure}[ht]
\centering
\includegraphics[width=0.48\textwidth]{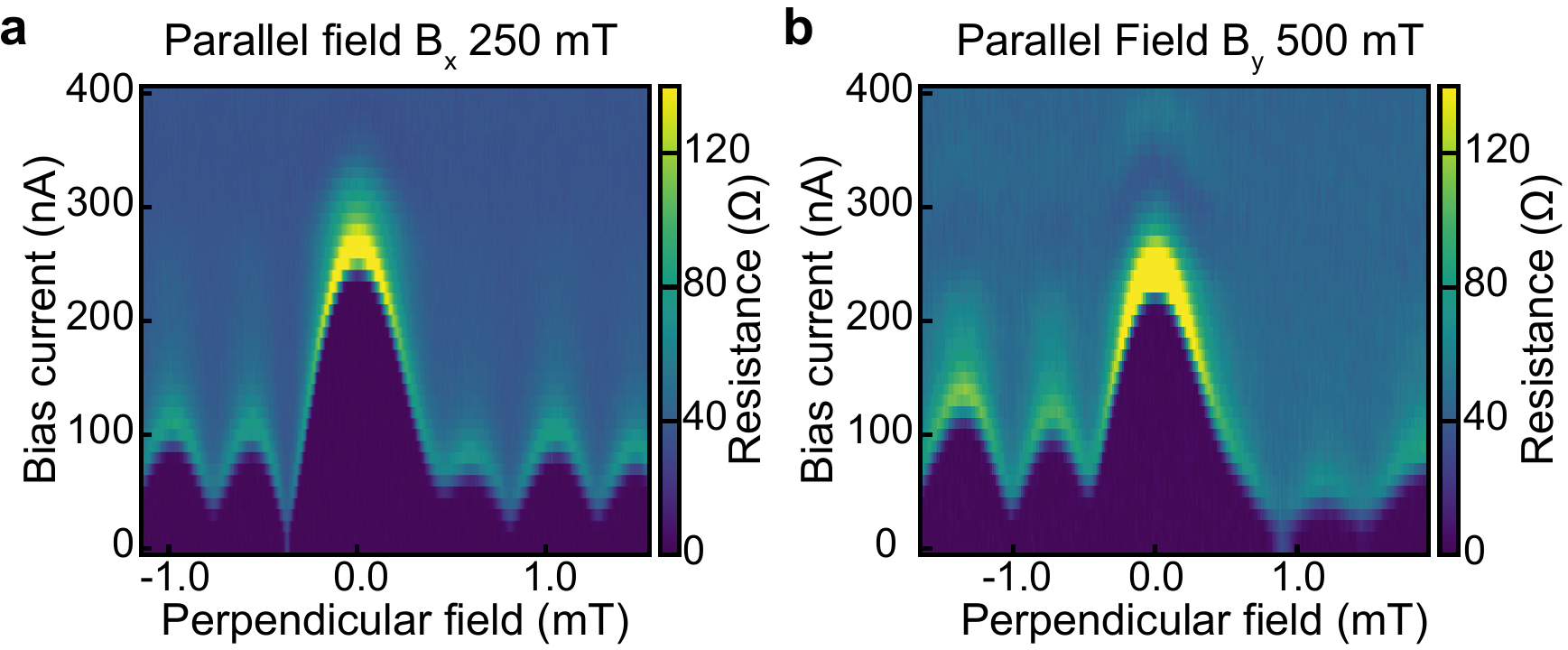}
\caption{\label{figure_s1}(Color online) Fraunhofer pattern of JJ 1 in the presence of an in-plane field ($V_g^1 = 0 V$, $V_g^2 = -7 V$). a) Fraunhofer pattern when applying 250 mT along the x direction i.e. parallel to the current. b) Fraunhofer pattern when applying 500 mT along the y direction. }
\end{figure}

The application of an in-plane magnetic field on the sample leads to a reduction of the critical current of the Josephson and a distortion of the Fraunhoffer pattern as illustrated in Fig \ref{figure_s1}.

The change in the critical current of the junction appears to strongly depends on the direction of the applied in-plane field. In Fig \ref{figure_s1}, the amplitude of the critical current is similar in both plots but the magnitude of the applied magnetic field is twice as large in the y direction compared to the x direction. 

For both directions of the field, the Fraunhoffer pattern appears asymmetric which is not the case in the absence of the in-plane as illustrated in the main text. The observed distortions are similar for both orientations of the field. 

When comparing those data to the ones presented in the main text, one can notice that the width of the first node has been divided by about two. We attribute this effect, which is also visible in the SQUID oscillations, to the transition out of the superconducting state of the indium layer at the back of the sample. The transition occurs around 30 mT and does not impact our study otherwise.

\section{Impact of the junction transparency on the SQUID critical current}

The current phase relationship (CPR) of a Josephson junction with a high transparency present a notable saw-tooth like profile which leads to distortions of the typical SQUID oscillations. In the following we discuss how this affects our measurements.

\begin{figure}[ht]
\centering
\includegraphics[width=0.48\textwidth]{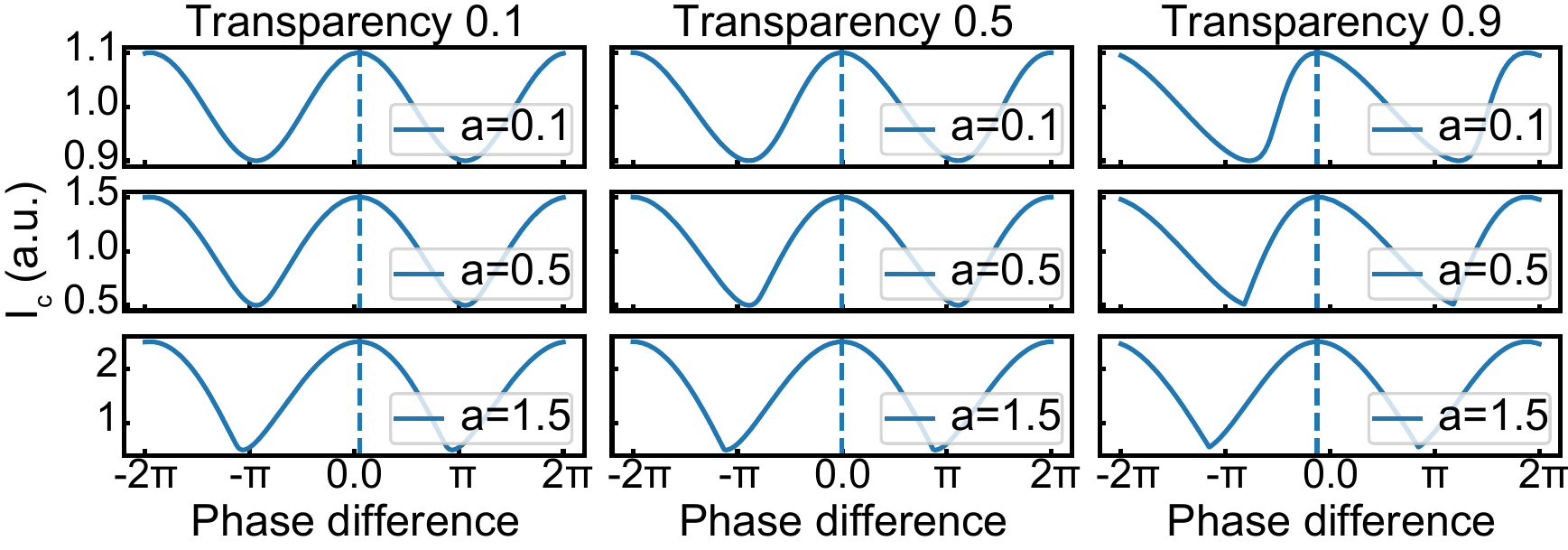}
\caption{\label{figure_s2}(Color online) SQUID critical current for highly transparent junction. The critical current of one of the junction is fixed to 1 and its transparency is set to 0.5. The values used for the other junction are the ones indicated on the figure. The method of calculation of the plotted current is the same one used to fit the experimental data. The dashed lines indicate the position of the maximum of the oscillation.}
\end{figure}

In Fig. \ref{figure_s2}, we present calculations performed for two junction of varying critical current and transparencies. For junctions with different transparencies, it appears that changing the relative amplitude of the current in each arm, $a = \frac{I_1}{I_2}$ of the SQUID does not alter the position of the maximum of the oscillation even though it can strongly alter the shape of the oscillation. This should not be surprising since the phase difference to be at the maximum of both CPR only depends on the shape of the CPR. This validates our method of extraction of the phase shift under the assumption that the applied gate voltage does not affect the junction transparency.

\begin{figure}[b]
\centering
\includegraphics[width=0.48\textwidth]{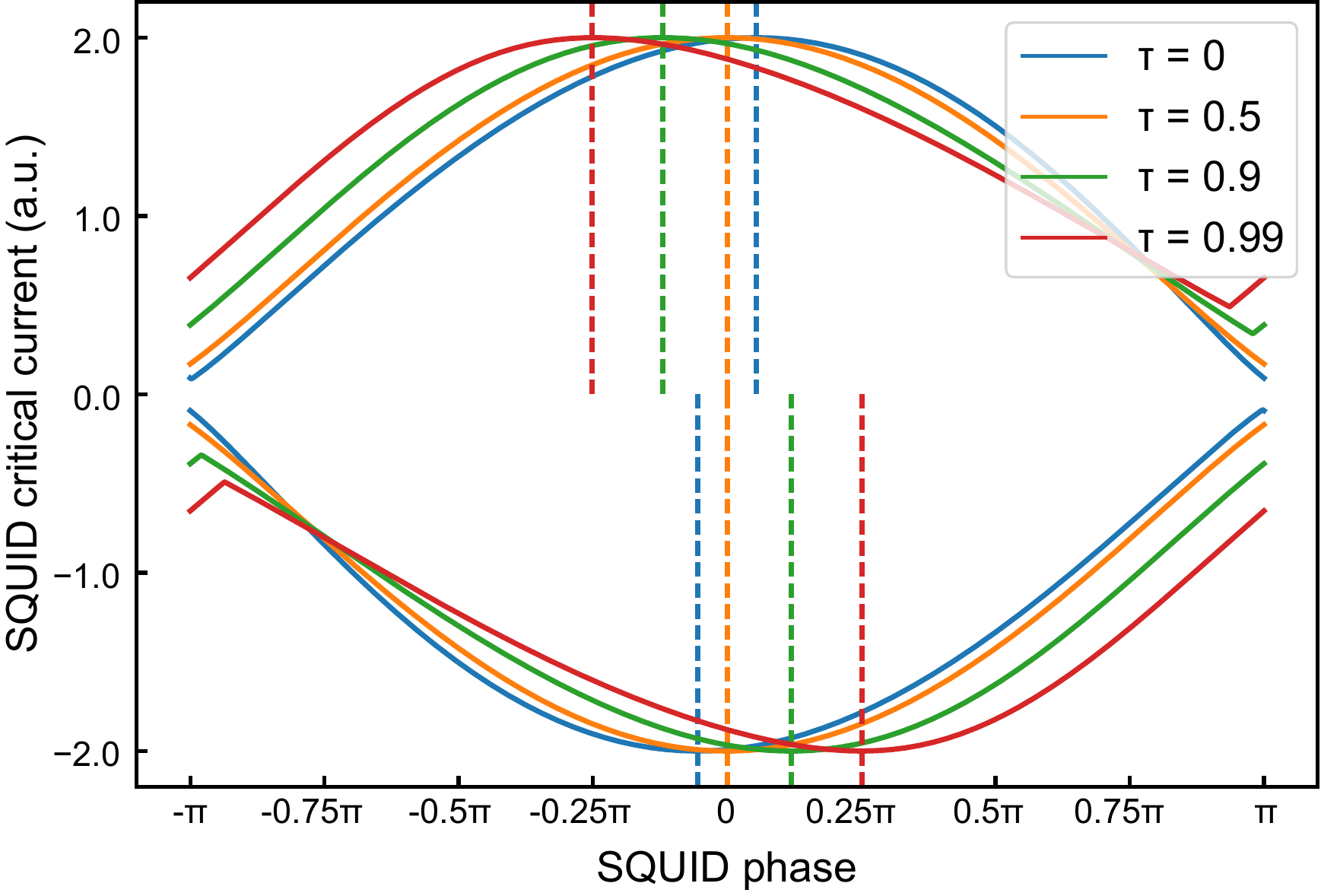}
\caption{\label{figure_s3}(Color online) SQUID critical current (positive/negative) for varying transparency of one junction. The transparency of the other junction is fixed at 0.5 and the current in both amplitudes are taken equal. The dashed lines indicate the position of the maximum/minimum of the oscillation.}
\end{figure}

In Fig. \ref{figure_s3} we illustrate the artificial phase-shift that can be induced by varying the transparency of one junction while the other is kept at a fixed transparency (0.5). We consider equal current in each arm but as mentioned above this has no consequence on the phase-shift. As the transparency is varied between 0 and 0.99, the oscillations are shifted by about $0.25\pi$ which is about half of the largest phase-shift we measured. Furthermore that shift has the opposite sign on the positive and negative branchs of the SQUID critical current which allows us to rule out this effects as being the dominant mechanism in our experiment as illustrated in Fig. 3 of the main text.

To reduce the measurement time, we have often worked with only the positive branch of the SQUID critical current and assumed a constant transparency of the junction as a function of the gate. This can lead to errors in the determination of the phase-shift obviously but as discussed above we have checked that a varying transparency cannot alone explain all our results.

\section{Fits used to prepare Fig. 4}

\begin{figure}[hb!]
\centering
\includegraphics[width=0.98\textwidth]{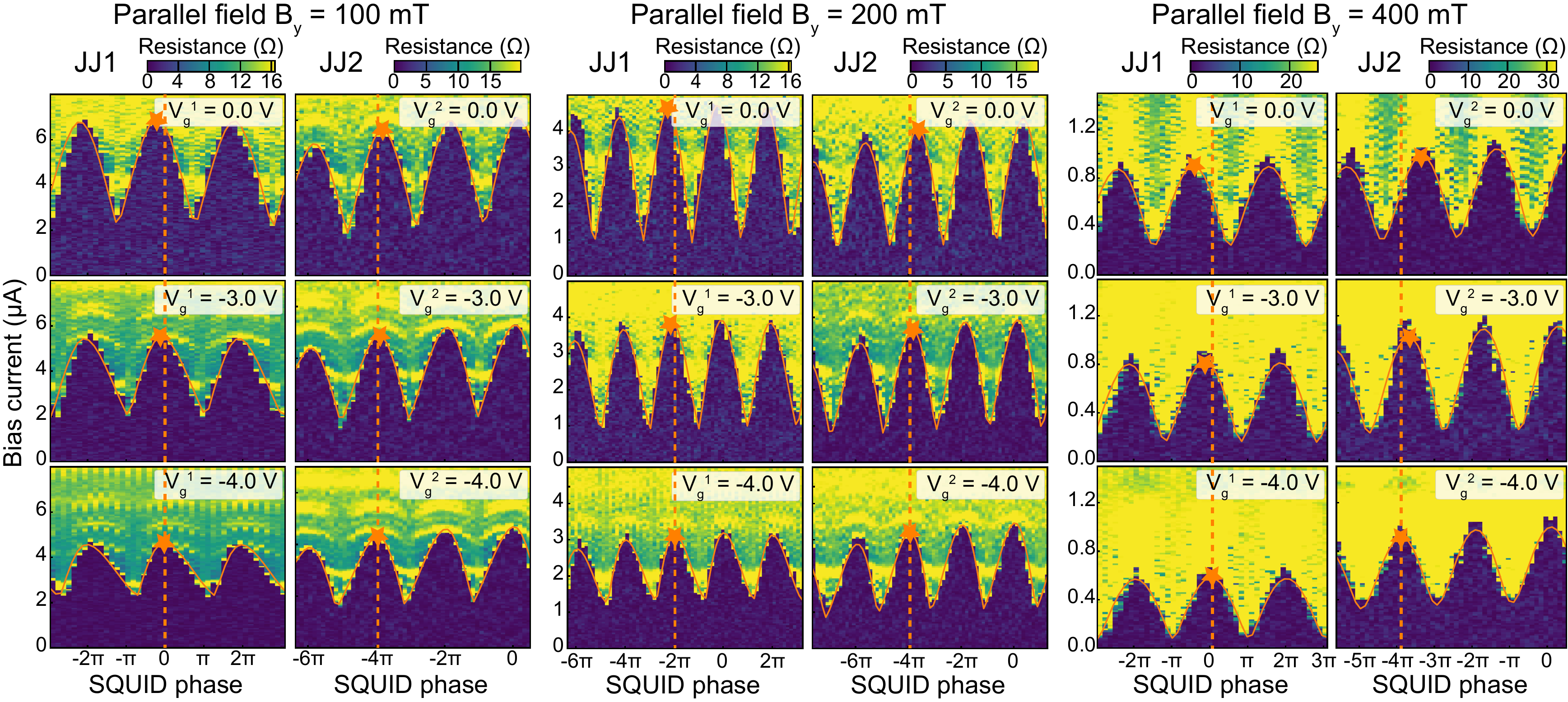}
\caption{\label{figure_s4}(Color online) Fits performed simultaneously (see Methods) on JJ1 and JJ2 data to extract the phase shift. When working on JJ1, Vg2 is set to 0 V, when working on JJ2, Vg1 is set to -2 V}
\end{figure}

The phase-shift of JJ2 as a function of the applied field presented in Fig. 4 of the main text has been extracted by fitting the SQUID oscillations of both JJ1 and JJ2 in a constrained manner as described in the Methods section of the main text. We present in Fig. \ref{figure_s4}, the data and fits obtained at three different values of magnetic field. As in the main text, we mark the position of the maximum at Vg = -4 V using a dashed line and the position of the maximum at each field using a star.

One can observe that the phase-shift observed for JJ1 is of the same order of magnitude than the one for JJ2 but of the opposite sign as expected from the SQUID equation.

\section{Phase-shift dependence on the magnetic field orientation}

According to most theoretical predictions \cite{Tokatly:2015,Konschelle:2015,Shen:2014},in the absence of Dresselhaus spin-orbit coupling applying a magnetic field along the x axis should not give rise to an anomalous phase. In InAs, the spin-orbit interaction is expected to be mostly of the Rashba type and we hence expect a reduction of the phase shift by rotating the field.

\begin{figure}[ht!]
\centering
\includegraphics[width=0.98\textwidth]{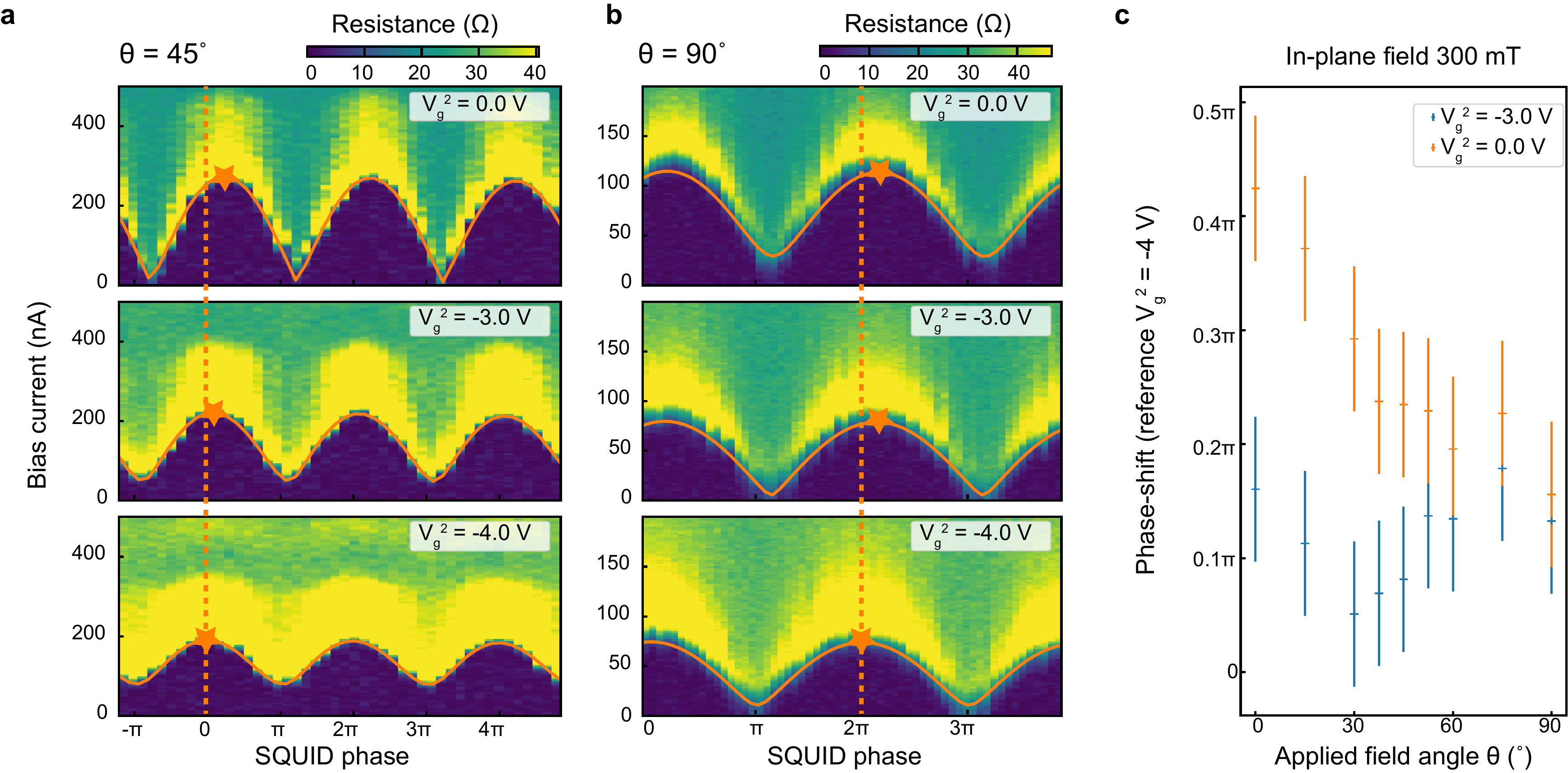}
\caption{\label{figure_s5}(Color online) JJ2 data and fits performed with an in-plane field of 300 mT applied at $\theta=$ 45˚(a)/90˚(b) with respect to the y-axis. (c) Phase-shift extracted from the fits as a function of $\theta$.}
\end{figure}

We present in Fig. \ref{figure_s5}, data taken in the presence of a 300 mT field at 45˚ (a) and along the x-axis (b) along with the extracted phase-shift as the function of the angle $\theta$ defined in Figure 1 c of the main text.

The phase-shift appears to diminish as we rotate the field away from the y-axis but remains finite as illustrated in (a) and (b). The error bars on the determination of the phase-shift are large due to fluctuations of the SQUID period inside the dataset (up to maximum of $10\%$) that forced us to treat it in two separate subsets.

\onecolumngrid

\twocolumngrid

\clearpage

\bibliographystyle{naturemag}
\bibliography {Ref_an_phase}

\end{document}